\documentclass{article}
\usepackage{spconf,amsmath,graphicx}
\usepackage{mathrsfs, url, subfig}
\usepackage[fontsize=9.25pt]{fontsize}
\usepackage{wrapfig}
\usepackage{setspace}
\usepackage{fancyhdr}

\title{Motif-centric representation learning for symbolic music}
\name{Yuxuan Wu$^{1,2}$, Roger B. Dannenberg$^{2}$, Gus Xia$^{1,3}$}
\address{$^1$ Music X Lab, MBZUAI, $^2$  Carnegie Mellon University, $^3$  NYU Shanghai}

\begin{document}

%
\maketitle
\thispagestyle{fancy}
\fancyhead{} 
\lfoot{\footnotesize{© 20XX IEEE. Personal use of this material is permitted. Permission from IEEE must be obtained for all other uses, in any current or future media, including reprinting/republishing this material for advertising or promotional purposes, creating new collective works, for resale or redistribution to servers or lists, or reuse of any copyrighted component of this work in other works.}}
\renewcommand{\headrulewidth}{0mm}

\begin{abstract}
Music motif, as a conceptual building block of composition, is crucial for music structure analysis and automatic composition. While human listeners can identify motifs easily, existing computational models fall short in representing motifs and their developments. The reason is that the nature of motifs is implicit, and the diversity of motif variations extends beyond simple repetitions and modulations. In this study, we aim to learn the implicit relationship between motifs and their variations via representation learning, using the Siamese network architecture and a pretraining and fine-tuning pipeline. A regularization-based method, VICReg, is adopted for pretraining, while contrastive learning is used for fine-tuning. Experimental results on a retrieval-based task show that these two methods complement each other, yielding an improvement of 12.6\% in the area under the precision-recall curve. Lastly, we visualize the acquired motif representations, offering an intuitive comprehension of the overall structure of a music piece. As far as we know, this work marks a noteworthy step forward in computational modeling of music motifs. We believe that this work lays the foundations for future applications of motifs in automatic music composition and music information retrieval.

\end{abstract}


\vspace{-0.2cm}
\section{Introduction}
\vspace{-0.2cm}
\label{sec:intro}

\begin{figure} [t!]
	\centering
    \includegraphics[width=0.48\textwidth]{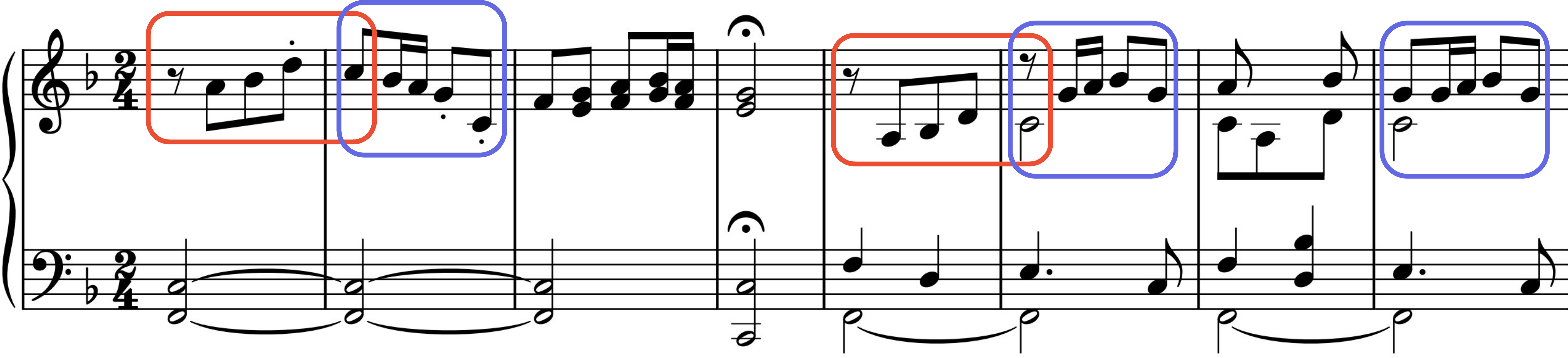}
    \vspace{-0.5cm}
	\caption{Motifs and their occurrences within a piece.}
	\label{fig:pastoral} 
 \vspace{-0.5cm}
\end{figure}

``Music is organized sound.'' The structured relationship among music elements is a significant factor in the meaning of music. Music
relies mostly on repetition and variation to create a sense of coherence, and the coherence can come from multiple levels of time scale. Among all the elements recurring in music, motifs are often the smallest meaningful part that, despite change and variation, is recognizable as present throughout \cite{schoenberg2006musical}. Motifs reappear multiple times in diverse forms, yet retain their recognizability and even iconic significance. As shown in Fig.~\ref{fig:pastoral}, the motif circled in red and the motif in blue develop their respective transformations as the music progresses. The coherence of motifs and variety in different occurrences of motifs endow music with the aesthetics of duality between constancy and changes, predictability and surprise, etc.

Understanding music motifs is pivotal for understanding the structure of music for machine music generation and high-level music information retrieval (MIR). Unfortunately, though it is easy for human listeners
to capture the coherence of motifs behind the changing appearances, it is usually a lot more difficult for computer programs, because musical language is more often implicit than straightforward. Rather than basic perspectives like pitch-based repetitions and modulations, effective motif analysis calls for the understanding of deeper and more abstract musical features.

In this paper, we propose a novel pretraining-based approach to acquire motif representations, and demonstrate its effectiveness through both theoretical analysis and experiments. We first introduce the training setting of motif-centric representation learning, and analyze the problem
with the classical contrastive learning method. To mitigate the problem, we introduce an alternative training method based on pure regularization and analyze its limitations as well. Based on the discussions, we propose a comprehensive approach that combines the advantages of both training methods. In the experiments, we incorporate a manually annotated dataset of motif labels for fine-tuning the pretrained models. Models are evaluated with a retrieval-based experiment, where the proposed method demonstrates a significant improvement over the baselines, with a 12.6\% increase in the area under the precision-recall curve (AUC-PR). Furthermore, we perform an intuitive visualization of music structure based on the learned representations, highlighting the potential of motif representations on downstream research applications. We hope this work can provide insights for automatic composition and MIR, as motif representations are crucial for analyzing and utilizing the coherence and diversity in compositions.


\vspace{-0.2cm}
\section{Related Work}
\vspace{-0.2cm}
One natural direction to study music motifs is finding a better measurement for symbolic music similarity. Due to the cognitive complexity of the task, finding a proper similarity measurement for symbolic music remains a challenge despite years of research \cite{velardo2016symbolic}. Traditional methods incorporate combinations of similarity functions, including edit distance, geometric distance, correlation coefficient, N-gram similarity measures, etc. \cite{suyoto2010simple, vempala2015empirically, janssen2017finding, park2019cross}. Recent neural network-based approaches have also shown great capabilities, particularly in robustness \cite{hirai2019melody2vec, karsdorp2019learning}.
Another closely related area is music pattern discovery. A classical general approach is the geometrical-inspired method, attempting to identify groups of notes with similar geometric features and extract these groups as musical patterns \cite{meredith2013cosiatec, Collins2013SIARCTCFPIP, bjorklund2022siatec}. Other rule-based methods utilize theoretical composition rules or statistical properties of data, demonstrating favorable outcomes \cite{nieto2012perceptual, lartillot2015automated, pesek2017symchm}. Despite abundant research, the assumptions involved in different approaches may not integrate well with each other, which to some extent hinders broader application. Similar issues apply to direct motif discovery attempts \cite{pinto2010relational, lartillot2014depth, benammar2017discovering}.

On the other hand, efforts have been made to model various aspects of music from the perspective of representation learning. For example, the representations of chords are well studied to resolve the problem of automatic chord recognition \cite{mcfee2017structured, jiang2019large, chen2021attend, 10099263}; other work incorporates proper inductive biases to disentangle composition style and musical content during style transfer, acquiring valuable representations for styles \cite{yangdeep, wang2020pianotree, wang2020learning}; other widely studied music representations include melody contours \cite{dai2021controllable, zou2022melons}, rhythm patterns \cite{lattner2019high, wei2019generating} and so on. However, modeling local musical information and structures remains underexplored \cite{dai2022missing, shih2022theme}.

In this paper, we aim to advance representation learning for music motifs, which is a complex yet essential aspect of music. Compared to traditional methods for motif analysis, we reduce the impact of assumptions to enhance compatibility with downstream models. Compared to representation learning work focusing on other music aspects, our work offers a complementary viewpoint.

\vspace{-0.2cm}
\section{Methods} 
\vspace{-0.2cm}
The target of representation learning for music motifs is to train a potent neural network encoder to bring together occurrences of the same motif in the learned embedding space. The supervision signals take the form of similarity or dissimilarity labels, indicating whether pairs of samples are associated with the same motif. We employ the Siamese network architecture for its effectiveness in learning from pairwise similarity labels \cite{bromley1993signature}. 

\vspace{-0.3cm}
\subsection{Contrastive Learning}
\begin{figure}[t]
    \centering
    \includegraphics[width=0.46\textwidth]{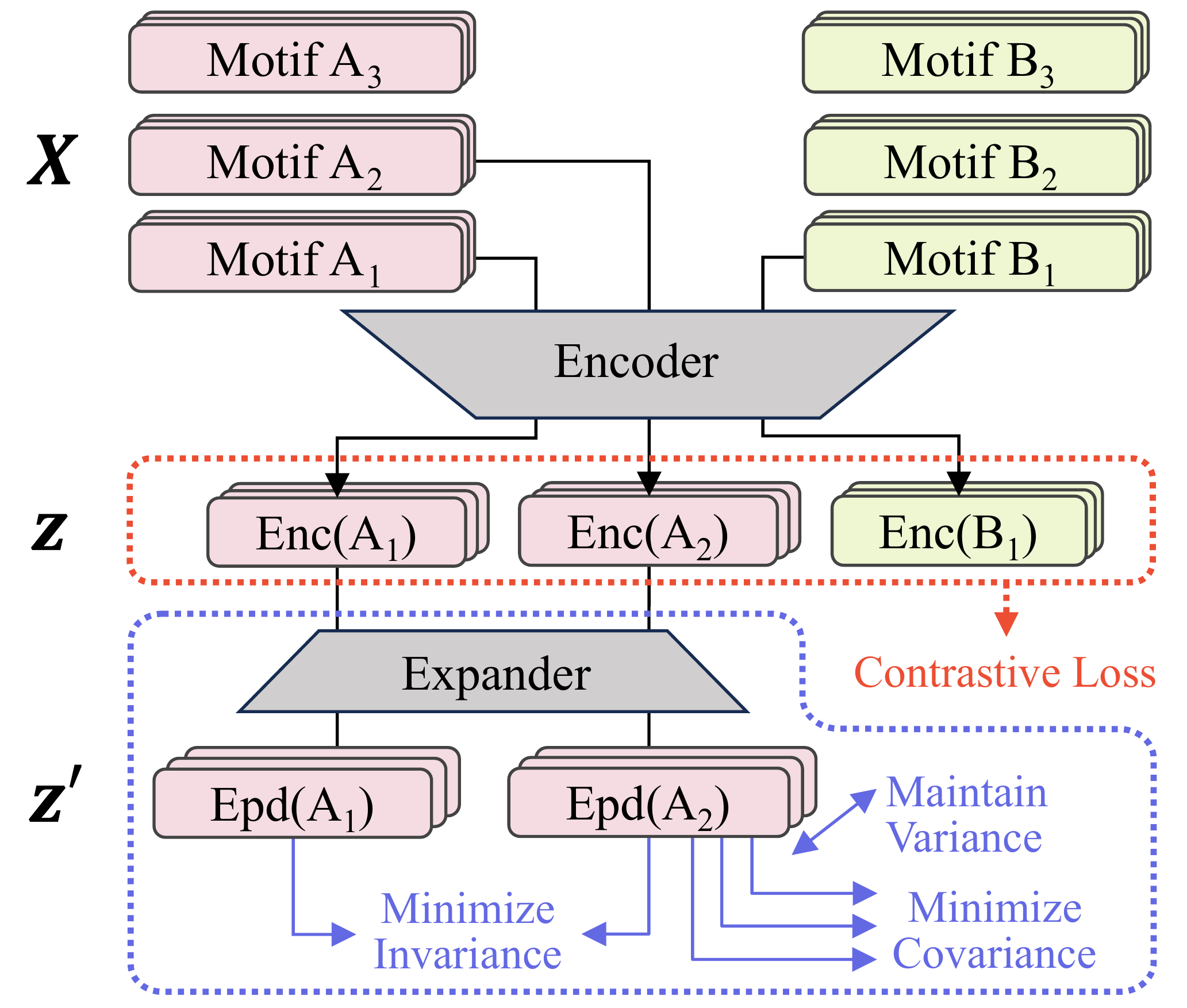}
    \vspace{-0.2cm}
    \caption{
        Architecture for music motif representation learning. Marked by red denotes the loss computation of contrastive learning. Marked by blue denotes the training process of VICReg.
    }
    \label{fig:arch}
    \vspace{-0.5cm}
\end{figure}

\vspace{-0.1cm}
\label{CL}
Contrastive learning is a classical method of SSL that harnesses the similarity and dissimilarity between data samples to learn meaningful representations without labeled data. In the context of motifs, for each anchor $\mathbf{X}$, its transformations serve as positive samples $\mathbf{X}^{+}$. Samples from other songs are considered negative samples $\mathbf{X}^{-}$. However, samples from the same song, while not necessarily distinct views of $\mathbf{X}$, are not considered negative samples due to potential motif repetitions within the same song, even if undiscovered \cite{shih2022theme}. The samples stochastically selected are passed through the encoder to obtain their respective embeddings $\mathbf{z}$, $\mathbf{z}^{+}$ and $\mathbf{z}^{-}$. The training objective is a contrastive loss function among three embeddings. In this study, we adopt triplet loss \cite{schroff2015facenet}, which can be formulated as:
\vspace{-0.2cm}
\begin{align*}
    \mathscr{L} = \max{(||\mathbf{z} - \textbf{z}^{+}|| - ||\mathbf{z} - \textbf{z}^{-}|| + margin, 0)}
\end{align*}
\vspace{-0.6cm}

where $||\cdot||$ is the distance metric, and $margin$ is a positive constant to prevent collapsing.
An illustrative training iteration is depicted in Fig.~\ref{fig:arch}, with $A_1$, $A_2$, $A_3$ as the occurrences of the same motif, and $B_1$, $B_2$, $B_3$ as the occurrences of another motif. The contrastive loss is computed with the positive pair of $A_1$ and $A_2$, and the negative pair of $A_1$ and $B_1$, as shown in the red box.

While contrastive learning is efficacious for acquiring a representation space in a self-supervised manner, it presents certain challenges when applied to music motifs. Although $\mathbf{X}^{+}$ is assuredly another view of $\mathbf{X}$, the randomly selected $\mathbf{X}^{-}$ might not be truly negative, as it is common that certain local music ideas can be shared across compositions of the same artist, genre or era, whether in the melody or accompaniment, and motifs are no exception. Considering similar segments from different pieces as distinct motifs is inaccurate and counterproductive for training. In fact, this noticeable adverse effect remains regardless of the data size. Approximating motif embeddings across all existing music by a Gaussian prior, the dataset can be seen as a random sample collection from this Gaussian. Given a fixed point $\mathbf{X}$, the expectation of the probability of selecting another point $\mathbf{X}^{-}$ from the dataset such that they share the same motif is equivalent to selecting such a $\mathbf{X}^{-}$ from the Gaussian.
Consequently, the expectation of the proportion of same-motif samples as negative pairs is invariant with the data size, and the adverse effect will not decrease with the introduction of more data. In the pretraining stage where the dataset is relatively large, the adverse effect should not be ignored.

\vspace{-0.3cm}
\subsection{Learning Through Regularization}
\vspace{-0.1cm}
\label{VICReg}

To address the problem introduced by contrastive learning, we suggest a regularization-based training method for music motif representation, drawing inspiration from VICReg in computer vision \cite{bardes2022vicreg}. This approach focuses on training with only reliable information - using only positive pairs while discarding negative ones. However, in contrastive learning, negative pairs not only serve as weak labels but also regularize the representation space against collapse. Without them, the loss function can converge to zero, yielding uniform embeddings for all inputs. Thus, proper regularization is vital when training solely with positive pairs. 

A post-processing module, known as the expander $Epd(\cdot)$, is attached to the encoder. $Epd(\cdot)$ takes in the embeddings $\mathbf{z} = Enc(\mathbf{X})$ and further projects them to the expanded embeddings $\mathbf{z}' = Epd(\mathbf{X})$ with a multi-layer perceptron. Training and regularization are achieved simultaneously by three loss terms, each playing a different role in training:

(1) The Invariance loss. This term helps the model learn the information in labels by pushing positive pairs closer. In a batch of $n$ samples, it can be computed as:
\vspace{-0.3cm}
\begin{align*}
    \mathscr{L}_{inv} = \frac{1}{n} \sum_{i=1}^n ||\mathbf{z'_i} - \mathbf{z'^+_i}||^2
\end{align*}
\vspace{-0.4cm}

(2) The Variance loss. This term aims to maintain sufficient variance within each batch, preventing the model from taking the shortcut that outputs a uniform value:
\vspace{-0.3cm}
\begin{align*}
    \mathscr{L}_{var} = \frac{1}{d'} \sum_{j=1}^{d'} 
    \max(0, 1-\sqrt{var([\mathbf{z'_{0}}_j, \hdots, \mathbf{z'_{n}}_j])})
\end{align*}
\vspace{-0.4cm}

where $d'$ denotes the dimension of the expanded embeddings, and $\mathbf{z'_{i}}_j$ denotes the $j$th dimension of $\mathbf{z'_i}$. The loss is implemented as the average from the $\mathbf{X}$ batch and $\mathbf{X^+}$ batch.

(3) The Covariance loss. This term encourages the off-diagonal values of the covariance matrix of the batched $\mathbf{z'}$ to be close to 0, decorrelating the dimensions of the expanded embeddings to ensure that they encode distinct information:
\vspace{-0.3cm}
\begin{align*}
    \mathscr{L}_{cov} = 
    \frac{1}{d'} \sum_{i\neq j} [Cov([\mathbf{z'_{0}}, \mathbf{z'_{1}}, \hdots, \mathbf{z'_{n}}])]_{i,j}^2 
\end{align*}
\vspace{-0.5cm}

The training process is shown in the blue-marked portion in Fig.~\ref{fig:arch}. Note that while the target of regularization is ultimately the $\mathbf{z}$ space, all loss terms are computed within the $\mathbf{z'}$ space. $Epd(\cdot)$ serves as a nonlinear projector, so that decorrelating the $\mathbf{z'}$ dimensions results in reducing the dependencies among $\mathbf{z}$ dimensions. Also, $Epd(\cdot)$ is only for regularization, and the final outcome is still the output of $Enc(\cdot)$. The final training objective is a weighted sum of $\mathscr{L}_{inv}$, $\mathscr{L}_{var}$ and $\mathscr{L}_{cov}$: 
\vspace{-0.2cm}
\begin{align*}
    \mathscr{L} = \alpha \mathscr{L}_{inv} + \beta \mathscr{L}_{var} + \gamma \mathscr{L}_{cov}
\end{align*}
\vspace{-0.6cm}

However, similar to contrastive learning, where false negative samples can harm the learning process, the VICReg-based approach also faces challenges. There is an underlying assumption in $\mathscr{L}_{var}$ that batch samples are generally dissimilar. While this assumption might hold when the data is abundant, it is quite questionable for smaller datasets, where the presence of same-motif samples in batches is highly probable. Attempting to preserve batch variance can increase the distance between occurrences of the same motif, counteracting the effect of $\mathscr{L}_{inv}$. As the dataset size increases, the likelihood of same-motif pairs appearing within the same batch decreases, consequently reducing the negative effects introduced by $\mathscr{L}_{var}$. This indicates the favorability of VICReg under large-data settings, but also points out its defect when the data is scarce. 

\vspace{-0.3cm}
\subsection{Pretraining and Fine-tuning}
\vspace{-0.1cm}
\label{functions}
Given the scarcity of music data with motif labels, it is ideal to pretrain the model to leverage the inherent structures within the larger-scale unlabeled data, and fine-tune the model on data with explicit motif labels.

In pretraining, the model learns via self-supervision on the auxiliary task of producing similar embeddings for different input views. To obtain different views of data samples, proper transformation functions need to be designed and applied as data augmentation. The principle behind these functions is changing the score while keeping the motivic idea to replicate motif variations in real music. Here we list some possible choices: (1) random transposition (2) random dropout (3) random note shift (4) random duration variation of the last note. The fact that each sample and its transformations are associated with the same motif provides natural weak labels for training. It should be noted that while it is impossible to exhaust all possible cases of motif variations with transformation functions, they aim to cover as many cases as possible while ensuring the credibility of the labels.

Based on the discussions in Sec.~\ref{CL} and \ref{VICReg}, we propose to combine the advantages of contrastive learning and VICReg. We use VICReg to pretrain the model with self-supervision, and turn to contrastive learning when fine-tuning the model on real data. This strategy allows the model to learn from reliable information in large data and make targeted refinements on smaller data.

\vspace{-0.2cm}
\section{Experiments}
\vspace{-0.3cm}
\subsection{Datasets}
\vspace{-0.1cm}

We conduct the study on pop arrangements, which feature clear motif boundaries, typical variations like transposition and inversion, and the polyphonic nature that covers the universality of music. We adopt the POP909 dataset for its clean arrangement data with beat and onset labels \cite{wang2020pop909}. Our model utilizes the Transformer encoder as its backbone, processing piano rolls through six Transformer encoder layers and an output module with a fully connected layer and pooling to generate embeddings.

For pretraining, all accompaniment tracks from POP909 are segmented into 1-bar chunks, which is a reasonable length for music motifs, especially for pop arrangements which are mainly driven by polyphonic patterns that carry the chords. Every chunk is processed 5 times with a random combination of the functions described in Sec.~\ref{functions}, resulting in a total of 6 views including the original. For fine-tuning, a carefully curated hand-labeled dataset is collected as an extension of POP909 \footnote[1]{\url{https://github.com/Irislucent/motif-encoder}.}. We use notes in an additional track to represent motif labels, covering occurrences of the same motif with the same pitch. The dataset comprises label tracks of 80 POP909 songs, including 20 songs for validation. We further process the raw data by matching every chunk to its corresponding motif label. Chunks with a label note covering over 75\% of their length are finally labeled with the note pitch. The finalized dataset has a mean number of motifs per song of 5.63, and a mean number of occurrences per motif of 7.58.

\vspace{-0.3cm}
\subsection{Model Training}
\vspace{-0.1cm}

To probe the potential problem of contrastive learning (CL), we make an adjustment in pretraining by filtering out samples with $>$50\% overlap with the anchor when selecting negative samples, which we refer to as negative-enhanced contrastive learning (NECL). Also, we include piano rolls pitch-shifted to start with the lowest notes as another baseline, which we refer to as interval-based piano rolls (IBPR). In fine-tuning, freezing the Transformer encoder layers and training only the output layers make only slight changes to the performance. Therefore, no model layers are frozen in the experiments in this section. 
We select the following empirical values: $margin=1$, $\alpha = 25$, $\beta = 25$ and $\gamma = 1$. All models are trained with AdamW \cite{loshchilov2018decoupled} with an initial learning rate of 1e-4 for pretraining, and 1e-5 for fine-tuning. 

\vspace{-0.3cm}
\subsection{Experimental results}
\vspace{-0.1cm}

\begin{figure*}[t]
    \centering
    \subfloat[]{\includegraphics[width=0.25\textwidth]{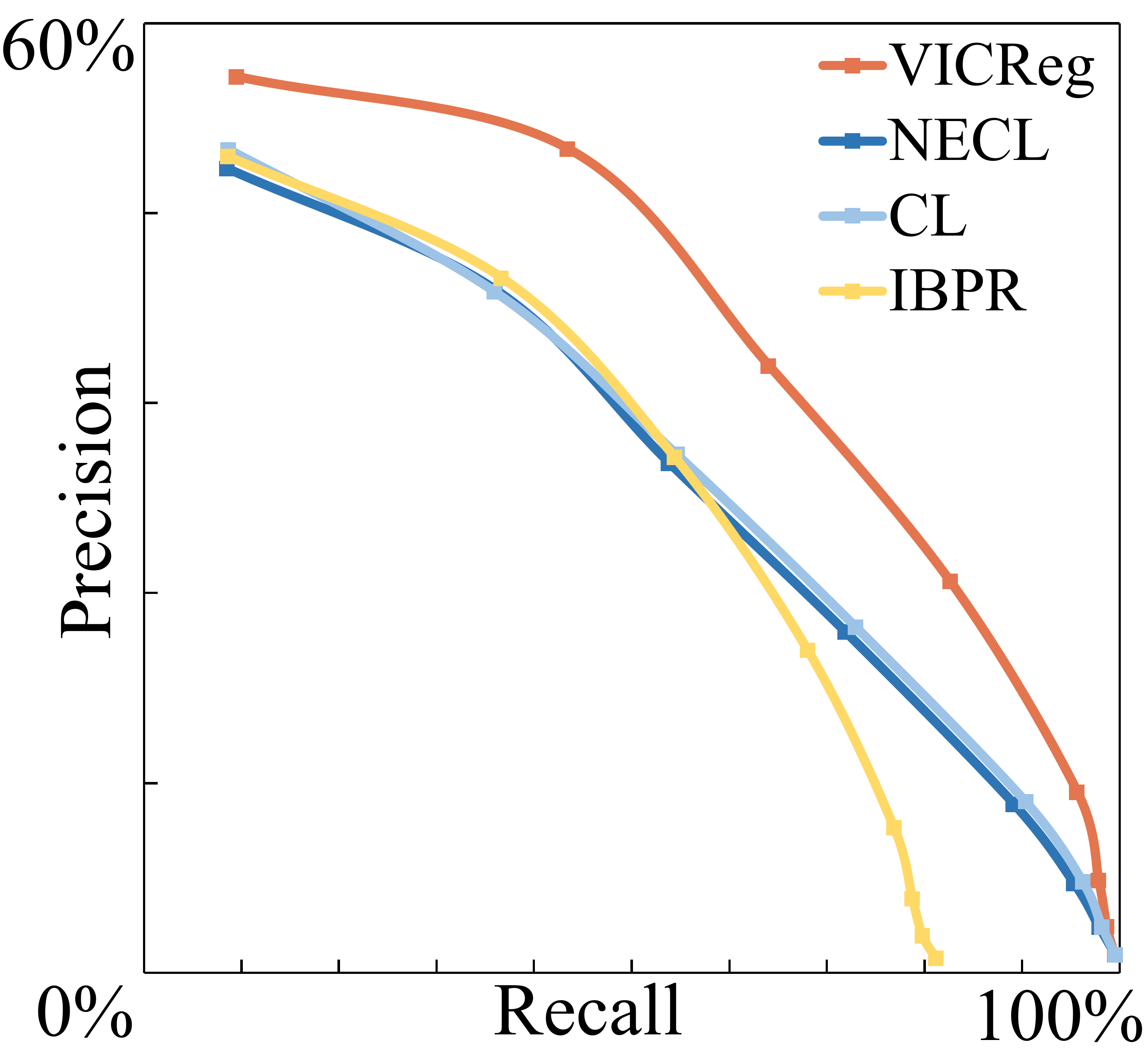}}
    \subfloat[]{\includegraphics[width=0.25\textwidth]{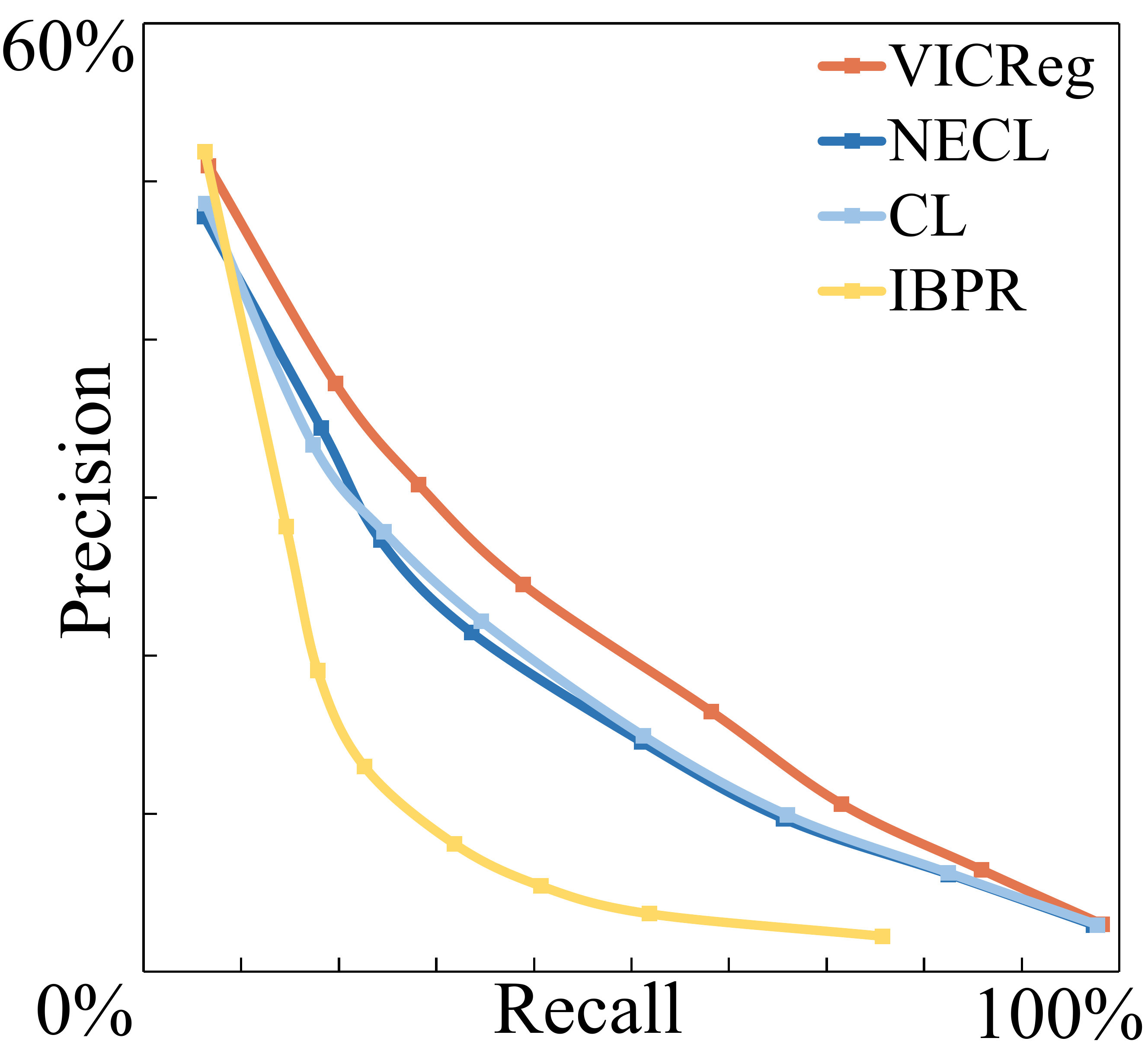}}
    \subfloat[]{\includegraphics[width=0.25\textwidth]{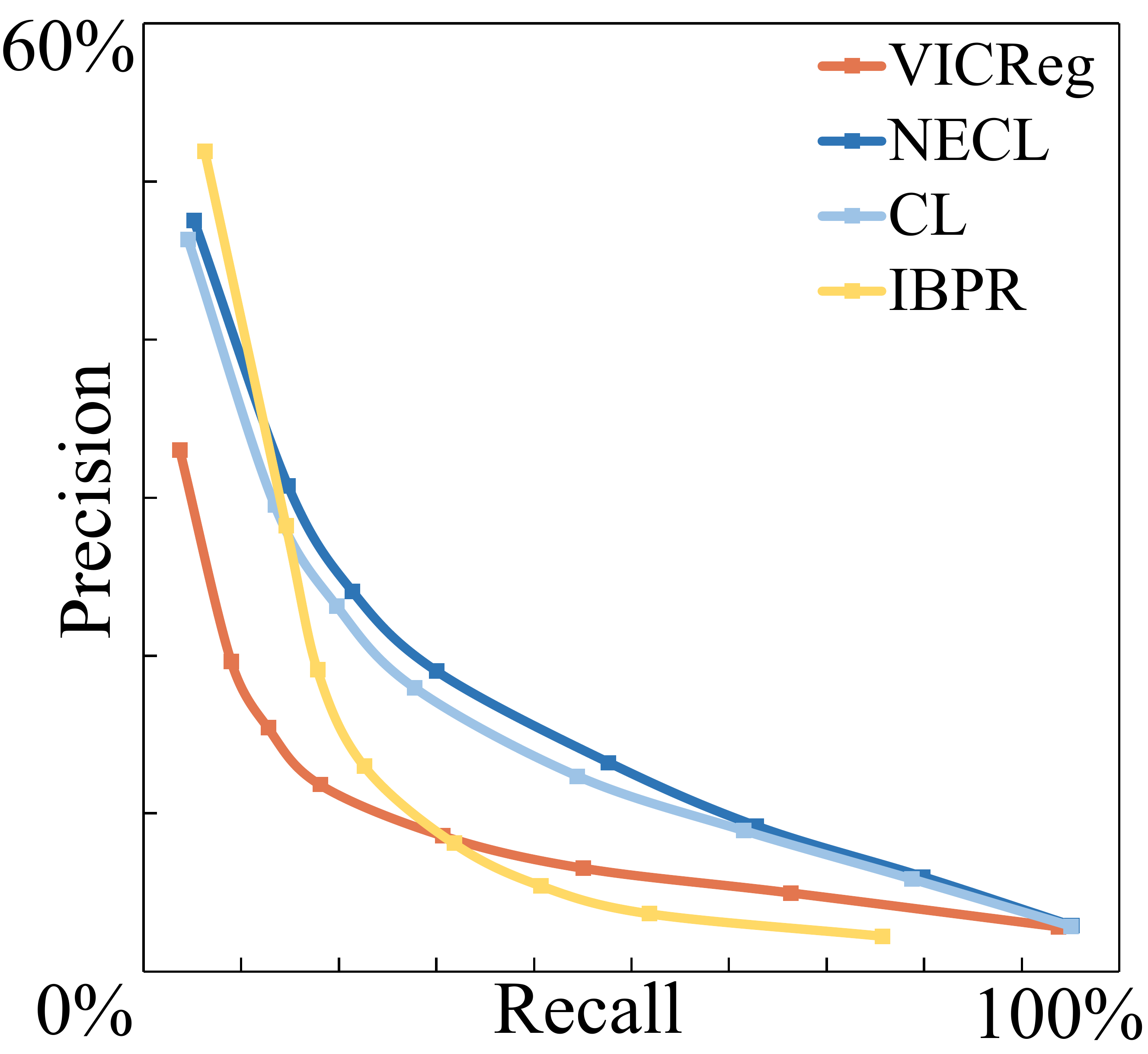}}
    \subfloat[]{\includegraphics[width=0.25\textwidth]{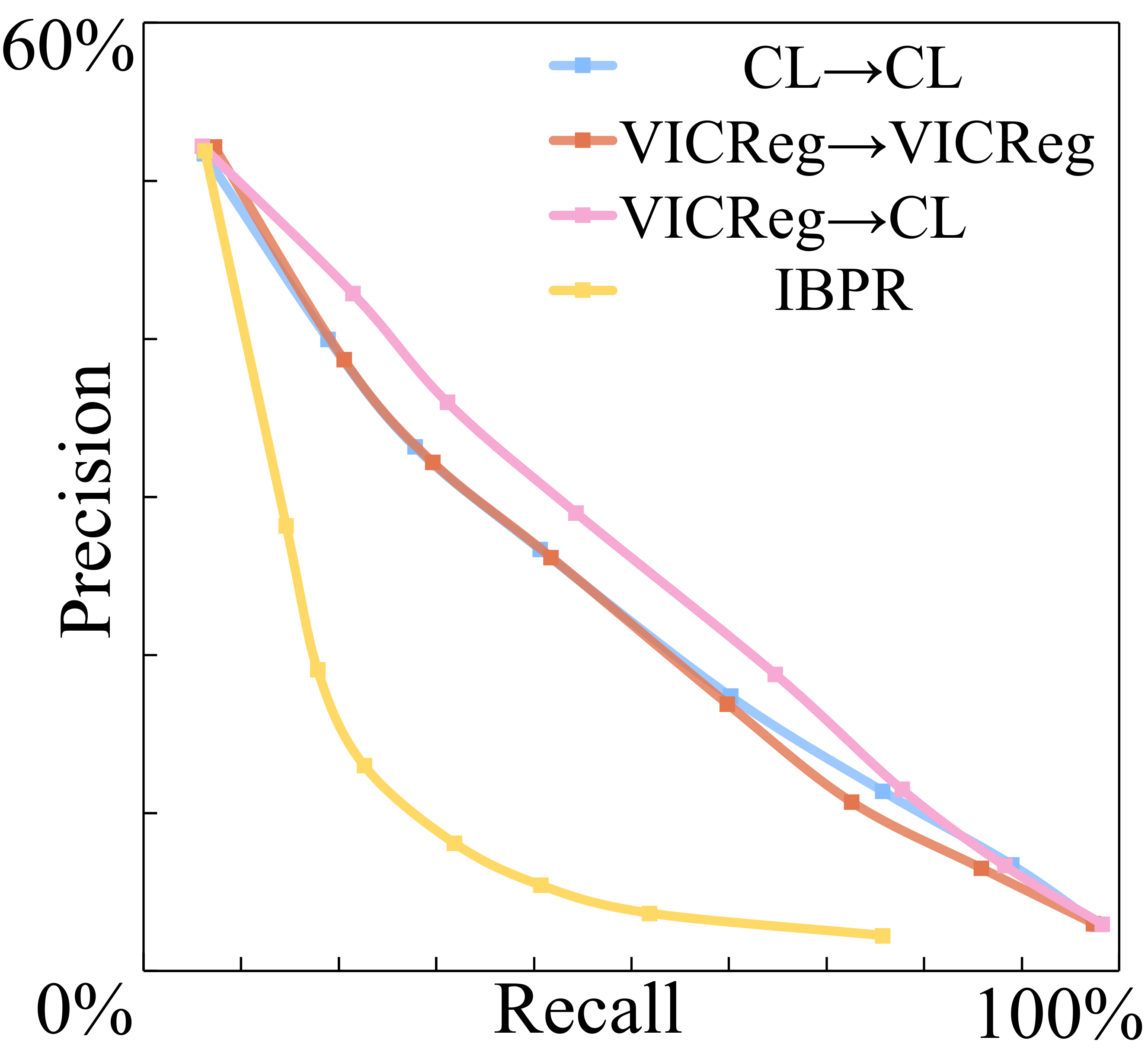}}
    \vspace{-0.3cm}
    \caption{PR curves of the evaluated methods. (a) Pretrained models on synthetic data generated by transformation functions. (b) Pretrained models on real data without fine-tuning. (c) Models trained from scratch on real data. (d) Pretrained models fine-tuned on real data.}
    \label{fig:curves}
    \vspace{-0.4cm}
\end{figure*}
 
The models are evaluated with a retrieval-based experiment. We compute the Euclidean distances from each anchor to every other sample in the representation space, and analyze the mean precision and recall of the top $K$ retrievals per anchor. For a comprehensive assessment, we conduct the following sets of experiments: 

(a) Evaluating pretrained models on synthetic data generated by the same transformation functions used in data augmentation.

(b) Evaluating pretrained models directly on real data.

(c) Evaluating models trained from scratch on real data.

(d) Evaluating pretrained models fine-tuned on real data, including the proposed method (VICReg$\rightarrow$CL), two direct fine-tuning baselines (VICReg$\rightarrow$VICReg and CL$\rightarrow$CL) and IBPR.

In each experiment, we draw precision-recall (PR) curves across various $K$ values. A curve closer to the top-right corner indicates better performance, signifying high precision and recall achieved concurrently. The results are reported in Fig.~\ref{fig:curves}.

In Fig.~\ref{fig:curves}(a), VICReg significantly excels other methods, which shows the effectiveness of leveraging only reliable information and proper regularization in self-supervision.  Fig.~\ref{fig:curves}(b) also demonstrates the better generalizability of VICReg compared to CL. Also, NECL does not show clear improvements compared to the vanilla CL, as false negative samples are still hard to eliminate, and picking only reliable negative samples will introduce sampling bias. However, VICReg falls far behind CL when trained from scratch on real data, as shown in Fig.~\ref{fig:curves}(c), which clearly illustrates the adverse effect brought by $\mathscr{L}_{var}$. In the fine-tuning experiment shown in Fig.~\ref{fig:curves}(d), the proposed method outperforms direct fine-tuning approaches by a noticeable margin, demonstrating the effectiveness of leveraging the respective advantages of CL and VICReg in different training stages. It is computed that the proposed method improves AUC-PR by 12.6\% compared to training with CL in both stages, and 13.4\% compared to training with VICReg in both stages. We also notice that IBPR generally falls behind deep learning-based methods, indicating the advantage of deep representations in terms of robustness.

\vspace{-0.2cm}
\section{Motif-centric Music Structure Visualization}
\vspace{-0.2cm}

We showcase the potential of motif representations by visualizing the music structure in pop piano arrangements from the perspective of motifs using the best-performing model from previous experiments. For demonstration, a random target song is parsed into chunks and encoded with the model. We adopt DBSCAN to cluster the embeddings, given that the number of motifs in music is unknown and noisy points are common \cite{ester1996density}. Embeddings in the same cluster represent occurrences of the same motif.

\begin{figure}[t]
    \centering
    \includegraphics[width=0.48\textwidth]{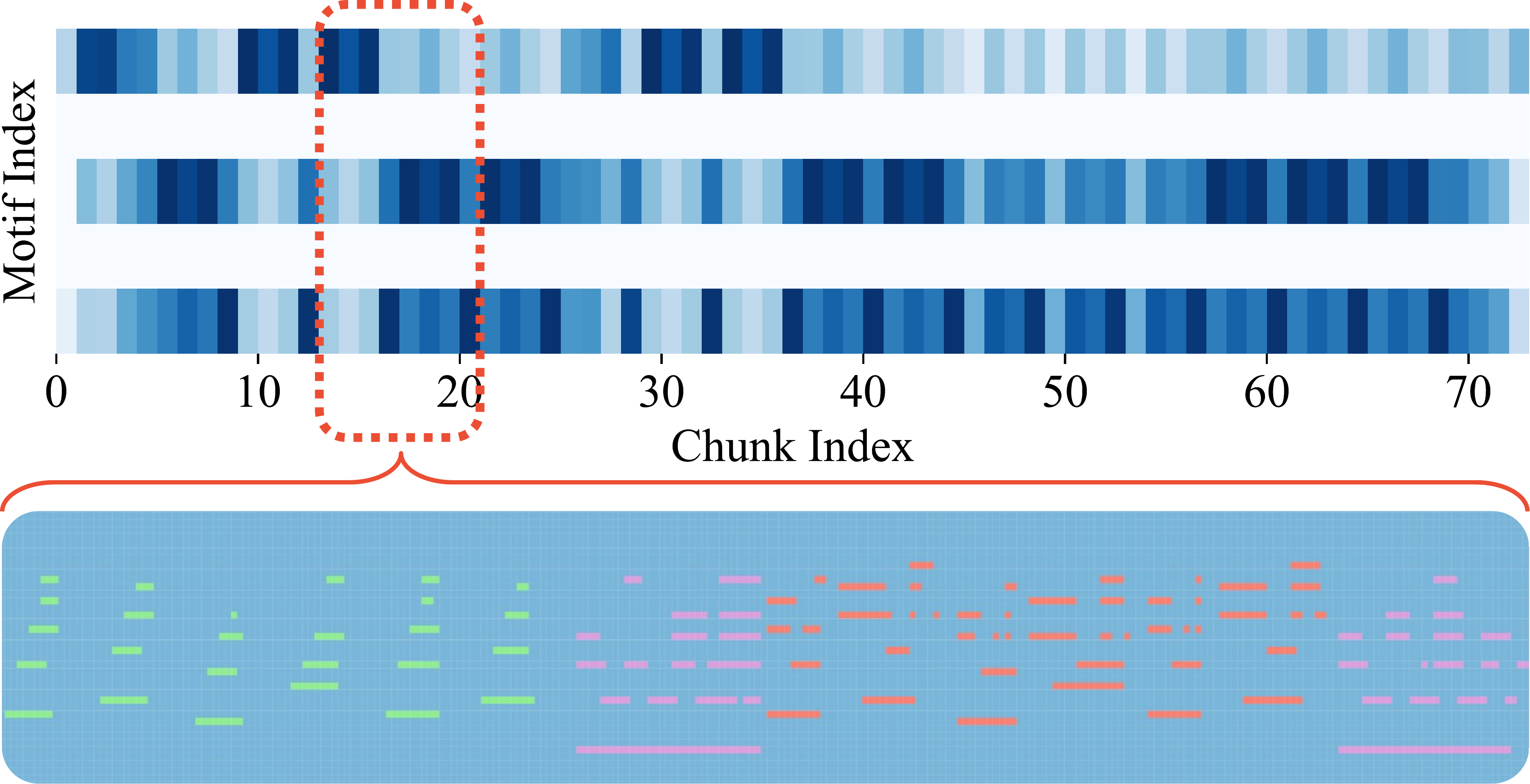}
    \vspace{-0.5cm}
    \caption{Motif-centric music structure visualization. Upper: The temporal structure of music in terms of motif distribution. Lower: Motif classifications in the music part marked by the red box.}
    \label{fig:vis}
    \vspace{-0.5cm}
\end{figure}

First, we visualize the temporal structure of music from the perspective of motifs by computing the distances from every embedding to every cluster center. We plot a heat map of the distance matrix to reveal the distribution of every motif over time, as shown in the upper part of Fig.~\ref{fig:vis}. Each row represents distance to a different motif. Second, we indicate the predicted motif labels of music segments by drawing the notes in the piano roll in different colors, shown in the lower part of Fig.~\ref{fig:vis}.

Fig.~\ref{fig:vis} reveals that the unique distribution of each motif effectively outlines the temporal structure of music. Some motifs tend to appear intensively at certain parts of the song. These can be motifs that appear very rarely in verses but shine prominently in the chorus. For example, the red motif in the piano roll drives the progression of the chorus, while the green motif constitutes the main part of the verse. Other motifs tend to emerge periodically in the song, possibly as certain transition motifs that play a role in every phrase, such as the purple motif in the piano roll. Meanwhile, variations among motif occurrences are noticeable. The model captures the underlying invariant motifs behind variations and differentiates musical segments according to their motifs.

\vspace{-0.2cm}
\section{Conclusions and discussions}
\vspace{-0.2cm}

This study has advanced motif-centric music representation learning via pretraining and fine-tuning methods. By combining the strengths of contrastive learning and VICReg, we have demonstrated noticeable improvements in representation quality. We have also illustrated the potential use of motif representations via music structure visualizations. 
Effective representations for motifs can serve as the cornerstone for downstream tasks in MIR and automatic composition, as motifs are often the entry point of music inspiration. For example, music generation algorithms could benefit from utilizing motifs for better structuredness and controllability. Also, the study of motifs could expand the scope of MIR and amplify its applicability. In conclusion, this study provides a foundation for deep learning applications in music motifs, contributing to the ongoing evolution of artificial intelligence in the realm of music. We hope our research can inspire further investigations, broadening the horizons of artificial intelligence and enriching people's lives with the companionship of music.



\newpage

\footnotesize
\bibliographystyle{IEEEbib}
\bibliography{strings,refs}

\end{document}